\documentclass[prl,twocolumn,showpacs,amsmath,amssymb,floatfix]{revtex4-1}
\usepackage{graphicx}  
\usepackage{dcolumn}   
\usepackage{bm}        
\usepackage[utf8]{inputenc}
\usepackage{graphics}
\usepackage{latexsym}
\usepackage{amsmath}
\usepackage{float}
\usepackage{pstricks,pst-plot}

\newcommand{\bra}[1]{\ensuremath{\langle #1 |}}
\newcommand{\ket}[1]{\ensuremath{| #1 \rangle}}

\renewcommand{\vec}[1]{\mathbf{#1}}

\newcommand{\ph}{\mathrm{ph}}
\newcommand{\ex}{\mathrm{ex}}
\newcommand{\pol}{\mathrm{pol}}
\newcommand{\phex}{\text{ph-ex}}

\newcommand{\exwf}{\psi_{\mathrm{ex}}^\tau}

\definecolor{mygreen}{rgb}{0.,0.55,0.}

\begin{document}


\title{Polariton Hall Effect in Transition-Metal Dichalcogenides}

\author{\'A. Guti\'errez-Rubio$^1$}
\author{L. Chirolli$^1$}
\author{L. Mart\'in-Moreno$^2$}
\author{F. J. Garc\'ia-Vidal$^{3, 4}$}
\author{F. Guinea$^{1,4,5}$}
\affiliation{$^1$IMDEA Nanoscience Institute, C/Faraday 9, E-28049 Madrid, Spain}
\affiliation{$^2$Departamento de F\'isica de la Materia Condensada, Instituto de Ciencia de Materiales, Universidad de Zaragoza, E-50009 Zaragoza, Spain}
\affiliation{$^3$Departamento de F\'isica Te\'orica de la Materia Condensada and Condensed Matter Physics Center (IFIMAC), Universidad Aut\'onoma de Madrid, E-8049, Spain}
\affiliation{$^4$Donostia International Physics Center (DICP), E-20018 Donostia/San Sebasti\'an, Spain}
\affiliation{$^5$School of Physics and Astronomy, University of Manchester, Manchester, M13 9PY, UK}

\date{\today}

\begin{abstract}
We analyze the properties of strongly coupled excitons and photons in systems made of semiconducting two-dimensional transition-metal dichalcogenides embedded in optical cavities. 
Through a detailed microscopic analysis of the coupling we unveil novel, highly tunable features of the spectrum, that result in polariton splitting and a breaking of light-matter selection rules. 
The dynamics of the composite polaritons is influenced by the Berry phase arising both from their constituents and from the confinement-enhanced coupling. 
We find that light-matter coupling emerges as a mechanism that enhances the Berry phase of polaritons well beyond that of its elementary constituents, paving the way to achieve a polariton Hall effect.
\end{abstract}

\maketitle


{\it Introduction.---}
The motion of composite excitations of quantum systems requires the introduction of gauge potentials related to their internal structure \cite{XCN10}. 
These potentials are generically defined in terms of the Berry curvature (BC) \cite{B84} in momentum space \cite{XCN10}. 
A generalization of these ideas to the energy bands of periodic systems has led to a deep understanding of the emergence of quantized topological properties and the associated appearance of edge modes \cite{TKKN82,NTW85}. 
Nontrivial Berry phases have been extensively discussed for many excitations, such as electrons, photons \cite{haldane2008possible} and excitons \cite{yao2008berry}, among others \cite{igarashi1988manifestation,SR16}, and the concept of Hall effect has been successfully extended to both photons \cite{onoda2004hall,hosten2008observation} and excitons \cite{estrecho2016visualising,onga2017exciton}.
Additionally, systems with strong light-matter interactions are being intensively studied, as they offer the possibility of modifying the material properties through coupling to light \cite{Hetal12}.  
Particular importance acquire two-dimensional materials, such as transition metal dichalcogenides (TMDs) \cite{Retal17}, embedded in optical cavities, as they represent an excellent platform where strong light-matter interactions can be studied \cite{wang2016coherent}.
At the same time, the band structure of semiconducting TMDs allows for nontrivial topological features.
The topological properties of excitons coupled to photons in quasi two-dimensional geometries have been first pointed out in Ref.~\cite{karzig2015topological}, where a winding phase in the coupling was recognized as the main ingredient to construct topological polaritonic crystals.

In this work we study strongly coupled excitons and photons in a single layer of $\mathrm{MoS_2}$ embedded in a cavity. 
We focus on the role of the composite Berry curvature, arising from both the bare constituents and their coupling, on the motion of the composite excitations, and the possibility to induce a polariton Hall effect. 
Through a detailed microscopic analysis we show that the cavity induces a breaking of polarization-valley locking \cite{zeng2012valley,mak2012control} that manifests in a cross coupling between right (left) circularly polarized photons and valley $K$ ($K'$) excitons, characterized by a winding phase.
This gives rise to a fine splitting of the upper (UPs) and lower polaritons (LPs).
We then analyze the impact of the winding coupling in the BC of the composite polaritons and show that it yields values for the hybrid modes far greater than those of the exciton and photon taken separately, showing that light-matter interaction dominates the 
BC in the strong coupling regime. 
By carrying out a semiclassical analysis of polaritonic wavepackets we estimate measurable topological Hall drifts, which may shed light to recent experiments, and pave the way to achieve a polariton Hall effect in TMDs.

{\it The model.---}
We consider excitons in a $\mathrm{MoS_2}$ monolayer embedded in an optical cavity. 
The maximum (minimum) of the valence (conduction) bands of a $\mathrm{MoS_2}$ monolayer lie at the two nonequivalent corners  of the hexagonal Brillouin Zone, $K$ and $K'$. 
The valence band shows a significant spin-orbit splitting, $\Delta_{\rm SO} \sim 100\,\mathrm{meV}$, resulting in a  spin and valley locking. 
Excitons in opposite valleys have opposite spin and are related by time-reversal symmetry. 
In the absence of a magnetic field, they are degenerate. 
Photons in perfectly conducting cavities are also doubly degenerate. 
This motivates a simplified model which includes two excitons, and the two photons closest in energy to the excitons. 
We assume perfect translational symmetry, and excitons and photons have a well defined parallel momentum, $\vec{\bf q}$.
For each value of $\vec{\bf q}$, the Hamiltonian evolution in the single-excitation subspace is described by a $4 \times 4$ matrix. 
For $\vec{\bf q} = 0$, this Hamiltonian can be further split into two $2 \times 2$ matrices, an electron-hole pair in a given valley can only be created by the absorption of a photon with a well defined circular polarization.
Away from $\vec{\bf q} = 0$ this selection rule \cite{zeng2012valley,mak2012control} does not apply, and the full $4 \times 4$ Hamiltonian needs to be considered.

We describe excitons with momentum $\vec{q}$ in valley $\tau$ through a variational Wannier wavefunction \cite{prada2015effective-mass}
\begin{align}
    \ket{ \exwf & (\vec{q}) }
    =
    \int d^2 q' \,
    \phi(q') \,
    [c_{\vec{q}'+\frac{\vec{q}}{2}}^{\tau}]^\dagger
    v_{\vec{q}'-\frac{\vec{q}}{2}}^{\tau}
    \ket{ 0 }
    \, .
    \notag
\end{align}
Here, $c_\vec{q}^\tau$ ($v_\vec{q}^\tau$) destroys a conduction (valence) electron in valley $\tau$ with wavevector ${\bf q}=(q_x,q_y)$, $\ket{0}$ denotes the filled Fermi sea, and $\phi(q)=\sqrt{\frac{2}{\pi}}a_{\rm ex}[1+(ka_{\rm ex})^2]^{-3/2}$ is the $s$-wave exciton wavefunction describing the relative motion of the bound electron and hole.
The variational parameter $a_{\ex}$ is the average radius of the exciton, which can be approximated by the typical value of $1\,\mathrm{nm}$ for $\mathrm{MoS}_2$ \cite{berkelbach2013theory,stier2016exciton}.
Our model therefore deals with Wannier excitons of small radii.
This point is relevant when studying 2D materials like $\mathrm{MoS}_2$, where dielectric screening changes significantly with respect to the 3D case.
In the continuum limit and defining $|\exwf (\vec{q}) \rangle = [b_{\vec{q}}^\tau]^{\dagger} |0\rangle$, the bare excitonic Hamiltonian reads
\begin{align}
    H_{\ex} & =
    \sum_{\tau}
    \int d^2q \,
    \left[
        \frac{ \hbar^2 q^2 }{ 2 M_{\ex} }
        + 2\Delta + E_b
    \right]
    [b_{\vec{q}}^\tau]^\dagger
    b_{\vec{q}}^\tau
    \, ,
    \label{eq:h_ex}
\end{align}
with an exciton mass $M_{\ex} = m_e + m_h \simeq 0.74~m_0$ ($m_e$ and $m_h$ being the masses of the bound electron and hole, and $m_0$ the electron rest mass), half gap $\Delta \simeq 1.5\,\mathrm{eV}$, and binding energy $E_b \simeq -1.1\,\mathrm{eV}$ \cite{rostami2015valley,komsa2012effects}.

Confined electromagnetic modes in a cavity with perfectly conducting mirrors and height $L_z$ have a momentum $\vec{k} = (\vec{q},k_z)$, with $\vec{q}=(q_x,q_y)$ and $k_z = \pi m/L_z$, with $m$ integer. 
We neglect the effect of fields associated to surface plasmons at the metallic boundaries of the cavity, as $L_z \simeq \lambda /2$, and also disregard modes with electric fields parallel to $\vec{e}_z$, as they do not couple to excitons in the $\mathrm{MoS_2}$ layer. 
In the basis of circularly polarized light, for which we define the bosonic operators $a_{\vec{q},k_z}^{\nu}$ with polarizations $\nu = \pm$, the photonic Hamiltonian reads (see Supplementary Material 
\footnote{Find the Supplemental Material at [reference to be provided by PRL].} for more details in the derivation)
\begin{align}
    H_{\ph} & =
    \sum_{\nu,k_z>0}
    \int d^2q \,
    \hbar \omega_{\vec{q},k_z}
    [a_{\vec{q},k_z}^\nu]^\dagger
    a_{\vec{q},k_z}^\nu
    \, ,
    \label{eq:h_ph}
\end{align}
where $\omega_{\vec{q},k_z}$ denotes the photon frequency.
As mentioned earlier, we focus on cavity sizes such that only the photon with $k_z = \pi/L_z$ interacts  strongly with the exciton, neglecting all the rest.
For simplicity we use $a_\vec{q}^\nu$ henceforth.

Considering in detail the microscopic coupling between excitons and photons (see Supplementary Material \cite{Note1}) the polaritonic Hamiltonian in the basis of circularly polarized photons $\{ b_{\vec{q}}^-, a_{\vec{q}}^+, b_{\vec{q}}^+, a_{\vec{q}}^- \}$ takes the form 
\begin{equation}
    \label{Ham4x4}
    {\cal H}=
    \left[
        \begin{array}{cccc}
            E_{\ex}
            & i \gamma
            & 0 
            & i \Gamma e^{2i\varphi}  \\
            -i \gamma
            & E_{\ph}
            & i \Gamma e^{2i\varphi}
            & 0                       \\
            0
            & -i \Gamma e^{-2i\varphi}
            & E_{\ex}
            & -i \gamma                \\
            -i \Gamma e^{-2i\varphi}
            & 0
            & i \gamma
            & E_{\ph}
        \end{array}
    \right]
    \, ,
\end{equation}
where $E_{\ex}$ and $E_{\ph}$ are the bare exciton and photon energies, respectively, and the couplings read
\begin{subequations}
    \begin{align}
        &
        \gamma = \gamma_0 \cos^2(\theta/2) 
        \, ,
        \label{eq:gamma}
        \\
        &
        \Gamma = \Gamma_{\ex}
        + \gamma_0 \sin^2(\theta/2) \, .
        \label{eq:Gamma}
    \end{align}
    \label{eq:gammas}
\end{subequations}
Here, $\cos\theta = k_z/\sqrt{ k_z^2 + q^2 }$ encodes the cavity size and
\begin{align} 
    & \gamma_0 = 
    \frac{ e \kappa \Delta }
         { \sqrt{ \pi\hbar\omega_{\vec{k}}
           \epsilon_0 L_z} }
    \left[
    F_0(\kappa)
    + F_1(\kappa) \frac{\hbar^2 q^2}{(\Delta/v_F)^2}
    \right]
    \, ,
    \notag
    \\
    & \Gamma_{\ex} = 
    \frac{ e \kappa \Delta }
         { \sqrt{ \pi\hbar\omega_{\vec{k}}
           \epsilon_0 L_z} }
    F_2(\kappa) \frac{\hbar^2 q^2}{(\Delta/v_F)^2}
    \notag
\end{align}
to order $\mathcal{O}[ (\hbar v_F q/\Delta)^2 ]$.
Moreover, $e$ is the electron charge, $\epsilon_0$ the vacuum permittivity and $\kappa = a_{\ex} \Delta /(\hbar v_F)$.
The real functions $F_0(\vec{\kappa})$, $F_1(\vec{\kappa})$ and $F_2(\vec{\kappa})$ are given in the Supplementary Material \cite{Note1}. 
The Hamiltonian Eq.~(\ref{Ham4x4}) contains the leading interactions in the system, although it neglects direct exciton coupling \cite{yu2014valley,yu2014dirac} and the dielectric character of the cavity, which would yield a TE-TM splitting \cite{panzarini1999exciton}.

The results in Eqs.~\eqref{eq:gammas} show that the strength of the couplings $\gamma$ and $\Gamma$ is tunable with the cavity width.
Whereas $\gamma$ is a mere renormalization of $\gamma_0$, $\Gamma$ contains two different contributions: the first, $\Gamma_{\ex}$, stems from the internal structure of excitons, particularly from the fact that they have a finite in-plane momentum; the second contains a term purely induced by the finite width of the cavity.
At this stage, a qualitative yet illustrative characterization of the system is at reach in terms of the parameters $\gamma$ and $\Gamma$.
The coupling $\gamma$ strongly hybridizes $\tau = +(-)$ excitons and $\nu = -(+)$ photons to yield four polariton bands, which are degenerate in pairs.
A nonzero $\Gamma$ lifts the degeneracy between both lower polariton (LP) and upper polariton (UP) at $q\neq 0$ and it breaks the valley-polarization selection rule.
Notice that the winding phases $e^{\pm i2\varphi}$ provide the required $\ell_z=2$ angular momentum that allows the coupling of a photon and an exciton with $\tau = \nu$.

It is convenient to compare the Hamiltonian Eq.~\eqref{Ham4x4} and the model in Ref.~\cite{karzig2015topological}.
In both models, the exciton-photon couplings show a winding phase and have the same asymptotic limits for both small and large $q$ values.
The calculation of the photon-exciton coupling, in our case, also takes into account the internal structure of the exciton. 
Furthermore, in Ref.~\cite{karzig2015topological} the use of a non-unitary transformation leads to a direct-photon coupling and an underestimation of the exciton-photon interaction in the strong-coupling regime ($q\sim k_z$) (see Supplementary Material \cite{Note1} for more details).

\begin{figure}
    \centering
    \includegraphics[width=0.45\textwidth]{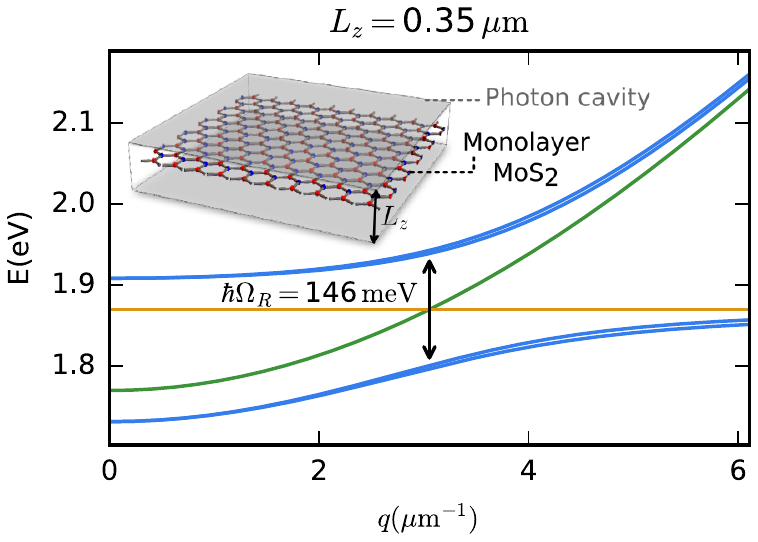}
    \caption{Dispersion relation of a bare exciton and a photon in a cavity and the two resulting polaritonic branches in the strong-coupling regime.}
    \label{fig:strong_coupling}
\end{figure}

In Fig.~\ref{fig:strong_coupling} we plot the dispersion relation for realistic values of the parameters of $\mathrm{MoS}_2$ excitons and photons in a microcavity of dimension $L_z = 0.35\,\mathrm{\mu m}$. 
In the strong-coupling regime one finds that $\gamma \sim 10^{-2}\,\mathrm{eV}$ and $\Gamma_{\ex}/\gamma_0 \sim 10^{-8}$. 
For these parameters $\Gamma_{\ex}$ can be neglected. 
However, it might be significantly enhanced in systems with a higher $\hbar v_F q/\Delta$ ratio, or greater exciton radii, such as Moir\'e patterns of graphene on top of an hBN substrate, where the gap ranges from 0 to $10\,\mathrm{meV}$ \cite{san-jose2014electronic}. 
On the other hand, we find that, within the strong-coupling regime,  $k_z/q \sim 1$. 
Thus, according to Eq.~\eqref{eq:gamma}, the finite size of the cavity results in a sizable $\Gamma$ and therefore plays a significant role in our system, as it is the principal source of selection-rule breaking.
Nevertheless, the fine splitting between the two upper polariton or lower polariton bands is probably too small to be measured experimentally 
\cite{wang2016coherent}. 
The predicted splitting might be observable in other systems that have a greater value for $\Gamma$.

{\it Berry Curvature.---}
In order to analyze the dynamics of the polariton, we need to include also the intrinsic Berry phases of the constituents, the exciton and the photon. 
The $n$th eigenstate of a generic single-excitation Hamiltonian such as Eq.~(\ref{Ham4x4}) takes the form $\ket{ \psi^n } = \sum_j \psi_j^n [\hat{\phi}^j]^\dagger \ket{0}$, with ${\cal H}\psi_n=E_n \psi_n$, $\hat{\phi}^j$ second quantized operators describing the constituents, and $|0\rangle$ the vacuum. 
When the state of the quasiparticle is not degenerate, we can study separately the Berry connection of each state, $\ket{\psi^n}$. 
The Berry connection is defined as $\vec{A}^n = i\bra{\psi^n} \vec{\nabla}_{\vec{q}} \ket{\psi^n}$. 
We define the Berry connection of the bare constituents as $\vec{A}_{0}^{ij} = i \bra{0} \hat{\phi}^i \vec{\nabla}_{\vec{q}} [\hat{\phi}^j]^\dagger \ket{0}$. 
These definitions lead to
\begin{equation}
    \vec{A}^n = 
    i[\psi^n]^\dag \nabla_{\vec{q}} \psi^n +
    [\psi^n]^\dag \vec{A}_{0} \psi^n \, .
    \label{eq:berry_connection}
\end{equation}
The Berry curvature can then be obtained as $\vec{\Omega}^n=\nabla_\vec{q}\times\vec{A}^n$.
The form of Eq.~\eqref{eq:berry_connection} is independent on the choice of the basis states defined by $\hat{\phi}^j$. 
This analysis agrees with the results of Ref.~\cite{yao2008berry}, where the BC of excitons, regarded as composite particles, is shown to be composed by an intrinsic term due to the BC of conduction and valence band electrons and an extrinsic one due to their coupling.

We first consider the intrinsic Berry curvature (IBC) of the exciton and the photon.
We neglect the dependence of $\phi(q)$ on the center-of-mass momentum and assume equal-mass conduction- and valence-band carriers.
Then, the exciton BC is \cite{yao2008berry}
\begin{align}
    \Omega_\ex^\tau(\vec{q})
    =
    \frac{1}{4}
    \sum_{\vec{q}'} |\phi(q')|^2
    \sum_{\beta=\pm 1}
    \Omega_c^\tau( \vec{q}' + \beta\vec{q}/2 )
    \, \nonumber,
    \label{eq:omega_ex}
\end{align}
where $\Omega_c^{\tau}$ is the BC of the bare conduction electrons (see Supplementary Material \cite{Note1}).
The Berry connection for circularly polarized photons reads $\vec{A}_\ph^\nu(\vec{k}) = \nu (\cos\theta - 1)\vec{e}_\phi$ and the resulting BC acquires the simple form $\Omega_\ph^\nu(\vec{k}) = \nu k_z/k^3$.

In the limit $\Gamma = 0$ the two polaritonic branches are doubly degenerate and we can label them with the valley index, $\tau$.
The UPs and LPs eigenstates are $|\psi_{\rm UP}^{\tau}\rangle = (u_q a_{-\tau,\vec{q}}^\dag + i\tau v_{q}b_{\tau,\vec{q}}^\dag) |0\rangle$ and $|\psi_{\rm LP}^{\tau}\rangle = (i\tau v_qa_{-\tau,\vec{q}}^\dag+u_{q}b_{\tau,\vec{q}}^\dag)|0\rangle$, with $u_q, v_q$ real normalized amplitudes.
Remarkably, polaritons show a finite BC that is due to the intrinsic contribution of the constituents. 
For UPs, we have
\begin{equation}
    \Omega_{\rm UP}^{\tau} =
    u_q^2 \Omega_{\mathrm{ph}}^{-\tau} +
    v_q^2\Omega_\ex^\tau +
    (\vec{A}_{\ph}^{-\tau} 
   - \vec{A}_{\ex}^{\tau})\cdot\hat{z}\times
    \vec{\nabla}_{\vec{q}} u_q^2
    \, ,
    \label{Eq:BCsel}
\end{equation}
and analogously for LPs with $u_q$ and $v_q$ interchanged. 
Clearly the BC of the composite system is significantly enhanced by the coupling, as $\nabla_{\bf q}u_q^2$ peaks in the strong-coupling regime \footnote{It may seem that the BC as described by Eq.~(\ref{Eq:BCsel}) is not gauge invariant, in that the Berry connection can be defined up to the gradient of a scalar function.
However, as pointed out in the section {\it Berry Curvature}, a redefinition of the wavefunction of the constituents affects both their Berry connection and their coupling, in a way that the result is basis independent.}.  

For $\Gamma\neq 0$ the polariton branches split.
The new eigenfunctions hybridize excitons from the two valleys, and photons with opposite polarizations with equal amplitude, and the Berry curvature of each quasiparticle vanishes. 
For the extrinsic Berry curvature (EBC), an explicit calculation shows that it also vanishes for all $\vec{q}\neq 0$.
However, the UPs and LPs are degenerate at $\vec{q}=0$, giving rise to a $\delta^{(2)}(\vec{q})$ structure of the BC.
A non zero Berry curvature arises if time reversal symmetry is broken, and a magnetic field induces a Zeeman coupling \cite{karzig2015topological,stier2016exciton}.
Alternatively, a polariton Hall current will exist when the initial state is given by a finite population of chiral excitations, which can be created using circularly polarized light.

\begin{figure}
    \centering
    \includegraphics[width=0.45\textwidth]{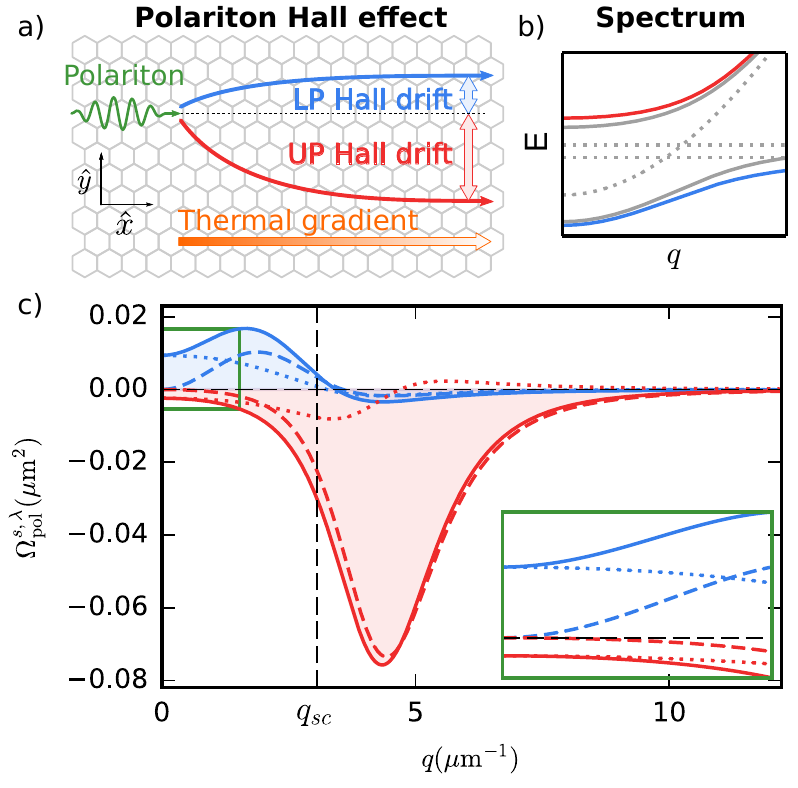}
    \caption{
    	a) Schematics of the Hall drift experienced by LP and UP upon application of a thermal gradient.
	    b) Zeeman split polariton spectrum.  
        c) Extrinsic (dashed), intrinsic (dotted) and total (solid) BC for the UP of highest (red) and the LP of lowest (blue) energy.
        Dashed black lines are a guide to the eye, with $q_{sc}$ such that $E_{\ex}(q_{sc}) = E_{\ph}(q_{sc})$.
        The lower inset zooms in the region of small momenta that appears squared in green.
    }
    \label{fig:bc}
\end{figure}

In the presence of a Zeeman coupling, $V_z$, the BC is smeared around small momenta.
A simple effective Hamiltonian for either the UP ($s=1$) or the LP ($s=-1$) branches can be obtained treating $\gamma$ and $\Gamma$ as perturbative parameters and considering $V_z \ll |E_{\mathrm{ph}}-E_{\mathrm{ex}}|$.
At small momenta,
\begin{equation}
    H^s_{\rm eff} = 
    f_s(q) +
    \left(
        \begin{array}{cc}
            \Delta_s(q)                 &
            \alpha q^2e^{2i\phi}      \\
            \alpha^* q^2 e^{-2i\phi}  &
            -\Delta_s(q)
        \end{array}
    \right) \, ,
\end{equation}
with $\alpha =-4\gamma\tilde{\Gamma} / (E_{\rm ex}-E_{\rm ph})$, $\tilde{\Gamma}=\Gamma/q^2$, and $f_s$ and $\Delta_s\propto V_z$ given in the Supplementary Material \cite{Note1}.
The correspondence between the polaritonic branches and gapped (gapless for $V_z=0$) bilayer graphene \cite{castro-neto2009electronic} becomes manifest here, providing with a qualitative understanding of the EBC.
The BC then reads
\begin{align}
    \Omega_\pol^{s,\lambda} (\vec{q}) =
    \frac{2 \lambda \Delta_s}{\alpha}
    \frac{q^2}
    {[q^2 + (\Delta_s/\alpha)^2]^{3/2}}
    \, .
    \notag
\end{align}
The BC for a generic parameter regime and $V_z = \gamma/5$ is shown in Fig.~\ref{fig:bc}  for one UP and one LP branch, the BC in the two other bands having the opposite sign.
A breakdown of the different terms in Eq.
\eqref{eq:berry_connection} is presented therein.
We can distinguish two different regions within the strong-coupling regime where either the intrinsic or the extrinsic contributions to $\Omega$ dominate.
For $q\to 0$, the EBC goes to zero whereas the IBC remains finite (see the inset in Fig.~\ref{fig:bc}).
In that range of momenta, the polariton behaves as merely inheriting the BC of its constituents.

Remarkably, it turns out that the BC of excitons is 6 orders of magnitude smaller than that of photons, and therefore can be neglected.
This fact allows the interpretation of the dotted curves in Fig.~\ref{fig:bc} in simple terms: a greater photonic component in the polaritonic eigenstate yields a higher value of the IBC.
This behavior combines with the decay of $\Omega_{\ph}$ with $q$ to make the IBC peak at $q=0$ for LPs and at $q>0$ for UPs.

As for the dominance of the EBC, it happens near the crossing between the bare exciton and the photon bands.
It becomes an order of magnitude greater than its counterpart for UPs, and around twice as much as the IBC for the LPs.
There is a significant difference in the BC between the UP and LP branches: the former reaches higher absolute values and peaks at larger momenta.
Once more, these features can be understood in terms of the competition between two factors.
Firstly, the proximity in energy between a pair of either UP or LP branches is expected to increase the value of the EBC.
Notice that UP dispersion relations come closer to each other with $q$, whereas the opposite happens with LPs, see Fig.~\ref{fig:bc}b.
Secondly, the EBC is expected to peak at momenta where the coupling is the greatest.
As a result of this balance, the location of the EBC peaks in Fig.~\ref{fig:bc}, which shift away from $q_{sc}$, can be explained.
We emphasized that photons are the only significant source of BC for polaritons via either the IBC or by means of the strong coupling with matter.

{\it Polariton Hall Effect.---}
The BC manifests in a Hall current perpendicular to an applied in-plane force.
On a general basis, assuming a polaritonic wavepacket in the $s$ branch and the $\lambda$ split band, centered at ${\bf r}_c = (x_c,y_c)$ and $\vec{q}_c = (q_{c,x},q_{c,y})$, and with energy $E^{s,\lambda}_{\pol}({\bf r}_c,\vec{q}_c)$, we can describe the evolution of its coordinates by semiclassical equations of motion that include the BC through an anomalous velocity \cite{chang1995berry,sinitsyn2008semiclassical}, 
\begin{equation}\label{eq:semiclassical}
    \dot{\vec{r}}_c =
    \frac{\partial
        E_{\pol}^{s,\lambda}}
        { \partial\vec{q}_c }
    -\dot{\vec{q}}_c \times
    \vec{\Omega}_\pol^{s,\lambda}(q_c) \,,
    \qquad
    \dot{\vec{q}}_c =
    -\frac{\partial
        E_{\pol}^{s,\lambda}}
        { \partial\vec{r}_c }.
\end{equation}
Due to the anomalous term, a polariton Hall effect naturally arises when hybrid modes are accelerated.
An anomalous exciton current was observed in Ref.~\cite{onga2017exciton}, where the nonzero $\dot{\vec{q}}_c$ was provided by a thermal gradient applied to the sample.
The same scheme can be applied to polaritons in the strong coupling regime.

We now consider a polaritonic wavepacket initially centered at $\vec{r}_c = \vec{q}_c = 0$ under the influence of an effective force in the $\hat{x}$ direction, Eqs.~\eqref{eq:semiclassical} predict for the Hall drift
\begin{align}
    y_c = \int_{0}^{q_c} dq_c' \,
    \Omega_{\pol}^{s,\lambda}(q_c') \, ,
    \label{eq:drift}
\end{align}
where the angular symmetry of $E_{\pol}^{s,\lambda}(\vec{r}_c,\vec{q}_c)$ has been taken into account.
Remarkably, the result applies also in the case of  relativistic corrections that arise due to the coupling with photons.
Assuming the domain of integration to exceed the strong-coupling regime, we extend the integration to infinity, so that the area under the curves of Fig.~\ref{fig:bc} gives a good approximation of the polariton drift. 
For realistic values of the parameters, we obtain $y_c \simeq 0.2\,\mathrm{\mu m}$ for UPs and $y_c \simeq 0.03\,\mathrm{\mu m}$ for LPs. 
Note that the IBC drives a drift that only amounts to $\sim 10^{-2}\,\mathrm{\mu m}$.

{\it Conclusion.---}
We have presented a microscopic treatment of the exciton-photon coupling in single layer of $\mathrm{MoS}_2$ embedded in a photonic cavity, focusing on polariton spectrum, selection-rules, and Berry curvature. 
Remarkably, we find that the cavity size promotes a selection-rule breaking, that appears as a valley-polarization cross coupling, characterized by a winding phase. 
The coupling results in a splitting of the polaritonic branches and a strong enhancement of the polariton BC, much beyond their constituent contributions. 
The polariton BC peaks in strong-coupling regime and gives rise to a polariton Hall effect, thus promoting 2D materials as a platform to study the topology of hybrid light-matter states.

{\it Acknowledgments.---}
A. G., L. M.-M. and F. G. acknowledge the European Commission under the Graphene Flagship, contract CNECTICT-604391.
L. C. and F. G. acknowledge funding from the European Union's Seventh Framework Programme (FP7/2007-2013) through the ERC Advanced Grant NOVGRAPHENE (GA No. 290846), L. C. acknowledges the Comunidad de Madrid through the grant MAD2D-CM, S2013/MIT-3007.
L. M.-M. and F. J. G.-V. acknowledge financial support by the Spanish MINECO under contract No. MAT2014-53432-C5.  

\bibliography{polaritons}

%


\clearpage
\onecolumngrid

\begingroup
\leftskip=0cm plus 0.5fil
\rightskip=0cm plus -0.5fil
\parfillskip=0cm plus 1fil
    \textbf{\large Supplemental Material for ``Polariton Hall Effect
   \newline in Transition-Metal Dichalcogenides''}
   \par
\endgroup
\vspace{0.5cm}

\setcounter{equation}{0}
\setcounter{figure}{0}
\setcounter{table}{0}
\setcounter{page}{1}
\makeatletter
\renewcommand{\theequation}{S\arabic{equation}}
\renewcommand{\thefigure}{S\arabic{figure}}

\twocolumngrid
\section{Eigenstates and topology of the Dirac Hamiltonian}
\label{App:Eigenstates}

The band structure of TMDCs allows to characterize electrons by their momentum close to the $K$ and $K'$ points, $\vec{q}=\tau {\bf K}+\delta\vec{q}$, with $\tau=\pm 1$.
The strong spin-orbit interaction present in TMDCs  splits the valley gap in a spin-locked way, resulting in a valley-spin Hall effect \cite{xiao2012coupled,mak2014valley}.
We then regard the actual spin as a mute degree of freedom and refer henceforth only to valley excitons that form in the smallest spin-resolved gap.

The electronic Hamiltonian in the $\tau$ valley reads
\begin{align}
    {\cal H} = 
    \hbar v_F(\tau q_x\sigma_x +q_y\sigma_y)
    +\Delta \sigma_z-\mu \,,
    \label{eq:Hdirac}
\end{align}
where $v_F$ is the Fermi velocity, $\vec{q} = (q_x,q_y)$ is the in-plane wave vector and $\sigma_i$ are Pauli matrices.
For simplicity we neglect conduction/valence band asymmetry.
The Hamiltonian of Eq.~(\ref{eq:Hdirac}) is a generic Dirac model describing the low-energy carriers in a number of materials, including surface states of topological insulators upon the identification of $\sigma_i$ with spin Pauli matrices \cite{garate2011excitons,efimkin2013resonant}, gapped graphene 
\cite{xiao2007valley-contrasting} and TMDCs \cite{xiao2012coupled}.
For $\mathrm{MoS}_2$, the values $\mu = 0.50\,\mathrm{eV}$, $v_F \simeq 6.5\cdot 10^5\,\mathrm{m/s}$ and $\Delta \simeq 1.5\,\mathrm{eV}$ were extracted from Refs.~\cite{rostami2015valley,komsa2012effects}.
The energy dispersions of the conduction and valence band read $\epsilon_{\rm c,v}=\pm \epsilon_{\bf k}=\pm \sqrt{\Delta^2+v_F^2k^2}$ and the associated eigenstates are 
\begin{align}
    & |\psi_c^\tau (\vec{q})\rangle
    =
    \left(
        \begin{array}{c}
            \cos(\Theta_\vec{q}/2)        \\
            e^{i\phi_{\vec{q}}^\tau}
            \sin(\Theta_\vec{q}/2)
        \end{array}
    \right) \,, 
    \\
    &
    |\psi_v^\tau (\vec{q})\rangle
    =
    \left(
        \begin{array}{c}
            e^{-i\phi_{\vec{q}}^\tau}
            \sin(\Theta_\vec{q}/2)        \\
            -\cos(\Theta_\vec{q}/2)
        \end{array}
    \right) \, ,
    \notag
\end{align}
where $\phi_{\vec{q}}^\tau={\rm arg}(\tau q_x+iq_y)$ and $\cos\Theta_\vec{q}=\Delta/\epsilon_\vec{q}$.
The Berry connection for the two valleys takes the form
\begin{align}
    \vec{A}_e^\tau (\vec{q}) = 
    \tau \sin^2(\Theta_\vec{q}/2)
    \vec{e}_\phi
    \, ,
    \label{eq:berry_connection_excitons}
\end{align}
and the single-particle BC for conduction (valence) electrons $\vec{\Omega}^\tau_{c(v)}(\vec{q})$ has a $z$ component \cite{xiao2007valley-contrasting,xiao2012coupled,zhou2015berry}
\begin{equation}
    \Omega_c^\tau = \Omega_v^\tau =
    \frac{ \tau \Delta (\hbar v_F)^2 }
    { 2 [\Delta^2 + (\hbar v_F q)^2 ]^{3/2} }
    \, .
\end{equation}

\section{Exciton-photon coupling}
\label{sec:ex_ph_coupling}

The general form for the vector potential inside a perfectly conductor cavity of width $L_z$ reads \cite{karzig2015topological}
\begin{align}
    \vec{A} & (\vec{r},t)
    =
    \sum_{\vec{q}, k_z > 0}
    {\cal A}_{q,k_z}
    e^{i(\vec{q}\cdot \vec{r}-\omega_{\vec{q}}t)}
    \Big\{
        \vec{e}_{\vec{q}_\perp}
        \sin(k_z z)
        \hat{a}_{\vec{q},k_z}^{\mathrm{TE}} 
        \notag
        \\
        &
        +
        \left[ 
            \vec{e}_{\vec{q}_\parallel} \sin( k_z z )
            \cos\theta
            -i \vec{e}_z \cos( k_z z ) \sin\theta
        \right]
        \hat{a}_{\vec{q},k_z}^{\mathrm{TM}}
    \Big\}
    \notag
    \\
    &
    + \sum_{\vec{q}}
    \frac{ {\cal A}_{q,0} }{ 2i }
    \vec{e}_z \hat{a}_{\vec{q},0}^{\mathrm{TM}}
    e^{i(\vec{q}\cdot \vec{r}-\omega_{\vec{q},0}t)}
    + \mathrm{h.c.} \, ,
    \label{Aperfectmetal}
\end{align}
where $ \vec{q} \in \mathbb{R}^2 $, $ k_z = m\pi/L_z $, $ m \in \mathbb{Z} $.
Here, TE and TM refer to transverse electric and transverse magnetic polarizations, $\hat{a}_{\vec{q},k_z}^{\mathrm{TE,TM}}$ to destruction operators of photons with momentum $(\vec{q},k_z)$, $\vec{e}_{\vec{q}_{\parallel (\perp)}}$ is a unitary vector parallel (perpendicular) to $\vec{q}$, and $\theta$ is the polar angle of the 3D momentum.
At last, 
\begin{align}
    {\cal A}_{q,k_z}
    & =
    i \sqrt{ \frac{ \hbar }
    { \epsilon_0 \mathcal{V}
    \omega_{\vec{q},k_z} } } \, ,
    \notag
\end{align}
where $\mathcal{V}$ is the volume of the system and $\omega_{\vec{q},k_z}$ the photon frequency.

Notice that the vector potential Eq.~(\ref{Aperfectmetal}) correctly describes cavities with perfectly conducting mirrors and does not lead to any TE-TM coupling in the bare cavity. 
It follows that two degenerate modes are present at every momentum ${\bf q}$. 
In contrast, a TE-TM coupling proportional to $(q_x\pm i q_y)^2$ arises in dielectric cavities, as originally shown in Ref.~\cite{panzarini1999exciton-light}.

\begin{figure}
    \centering
    \includegraphics[width=0.45\textwidth]{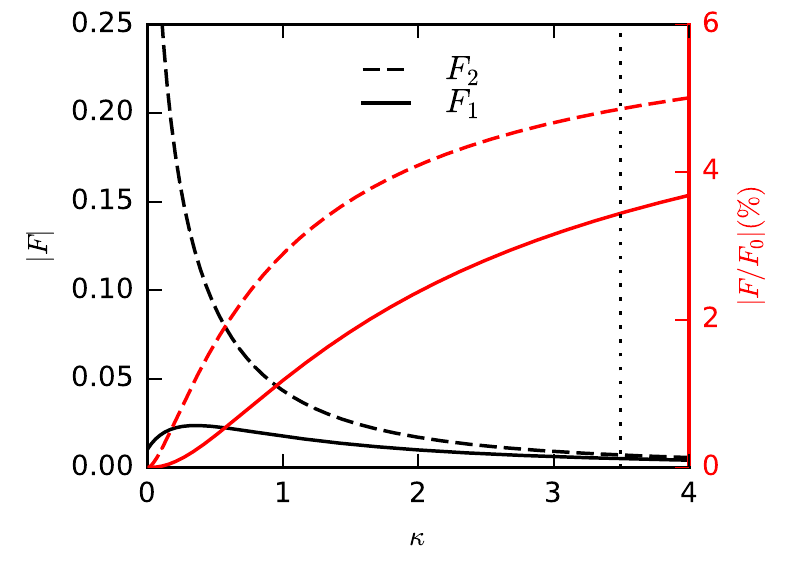}
    \caption{
        Plots of $F_{1}(\kappa)$ and $F_{2}(\kappa)$ and their quotient with $F_0(\kappa)$, which determine---together with the quantity $(\hbar v_F q/\Delta)^2$---the strength of the couplings induced by the finite momentum of the exciton.
        The vertical dotted line indicates the value of $\kappa$ for our model of excitons in $\mathrm{MoS}_2$.
}
    \label{fig:Fs}
\end{figure}

Polaritons arise when considering light-matter interactions.
Performing the standard minimal coupling substitution in the Dirac Hamiltonian of Eq.~\eqref{eq:Hdirac}, $\hbar{\bf k}\to\hbar{\bf k}-e\vec{A}$, one gets the electron-photon interaction \cite{san-jose2016inverse}
\begin{align}
    W_{\mathrm{em}} =
    -\frac{e}{\hbar}
    \int d^2r \,
    A_i(\vec{r})
    \psi_s^\dagger(\vec{r})
    [\partial_{q_i}\mathcal{H}]_{ss'}
    \psi_{s'}(\vec{r}) \, .
    \label{eq:w_em}
\end{align}
Here, $\psi_s^{(\dagger)}(\vec{r})$ is an electron destruction (creation) operator in real space, and the subindex $s$ refers to the basis considered in Eq.~\eqref{eq:Hdirac}.
Subsequently restricting the Hilbert space to free excitons and photons, one gets the Hamiltonian $H = H_{\ph} + H_{\ex} + H_{\phex}$, where the exciton-photon coupling reads
\begin{align}
    H_{\phex} =
    \sum_{\nu,\tau,k_z>0}
    \int d^2 q \,
    g_{\nu,\tau} (\vec{q},k_z)
    \,
    [a_{\vec{q},k_z}^\nu]^\dagger
    b_{\vec{q}}^\tau
    + \mathrm{h.c.}
    \label{eq:h_ph_ex}
\end{align}
The polarization- and valley-dependent couplings read
\begin{align} 
    g_{\mathrm{TE},\tau} (\vec{q}, k_z)
    & = 
    (\gamma_0 e^{-i\tau\varphi}
    -\Gamma_{\ex} e^{i3\tau\varphi})/\sqrt{2} \, ,
    \notag
    \\
    g_{\mathrm{TM},\tau} (\vec{q}, k_z)
    & = 
    i \tau \cos\theta
    (\gamma_0 e^{-i\tau\varphi}
    +\Gamma_{\ex} e^{i3\tau\varphi})/\sqrt{2} \, .
    \notag
\end{align}
Here, $\tan\varphi=q_y/q_x$, and $\gamma_0$ and $\Gamma_\ex$ were given in the main text,
\begin{subequations}
    \begin{align} 
        & \gamma_0 = 
        \frac{ e \kappa \Delta }
             { \sqrt{ \pi\hbar\omega_{\vec{k}}
               \epsilon_0 L_z} }
        \left[
        F_0(\kappa)
        + F_1(\kappa)
        \frac{\hbar^2 q^2}{(\Delta/v_F)^2}
        \right]
        \, ,
        \\
        & \Gamma_{\ex} = 
        \frac{ e \kappa \Delta }
             { \sqrt{ \pi\hbar\omega_{\vec{k}}
               \epsilon_0 L_z} }
        F_2(\kappa) \frac{\hbar^2 q^2}{(\Delta/v_F)^2}
        \, .
    \end{align}
    \label{eq:Fs}
\end{subequations}
to order $\mathcal{O}[ (\hbar v_F q/\Delta)^2 ]$.
Here, $e$ is the electron charge, $\epsilon_0$ the vacuum permittivity and $\kappa = a_{\ex} \Delta /(\hbar v_F)$.
The following definitions were also used:
\begin{align}
    &
    F_0(\kappa) =
    \frac{1}{\kappa}
    \left[ \frac{1}{\kappa} + \frac{1}{\kappa+1} \right] \, ,
    \notag
    \\
    &
    F_{1}(\kappa) =
    \frac{1}{8}
    \int_0^\infty dx \,
    x f_3(\kappa x) \times
    \notag
    \\
    & \quad
    \times
    \left[
        -\frac{1}{2} f_3(x) -\frac{1}{2}f_4(x)
        + x^2 f_5(x)
    \right] \, ,
    \notag
    \\
    & F_{2}(\kappa) =
    \frac{1}{8}
    \int_0^\infty dx \,
    x f_3(\kappa x)
    \Big[ 
        \frac{1-f_1(x)}{x^2} +
        \notag
        \\
        & \quad
        + \frac{1}{4}f_3(x) + \frac{1}{4}f_4(x)
        + \frac{1}{2} x^2 f_5(x)
    \Big] \, ,
    \notag
\end{align}
and $f_n(x) = (1+x^2)^{-n/2}$.
$F_{1}(\kappa)$ and $F_2(\kappa)$ can be expressed in terms of elementary functions, which is omitted here for the sake of simplicity.
Their plots and a comparison with the dominant term $F_0(\kappa)$ appear in Fig.~\ref{fig:Fs}.
As the expansion in $\hbar q/(\Delta v_F)$ suggests in Eqs.~\eqref{eq:Fs}, the source of both $F_1(\kappa)$ and $F_2(\kappa)$ is the $\mathcal{O}(q^2)$ term of the excitonic wavefunction, whereas $F_0(\kappa)$ can be traced back to the $\mathcal{O}(q^0)$ contribution.
We emphasize that the coupling $\Gamma_{\rm ex}$ properly accounts for the internal structure of the exciton, as the binding energy is not much smaller than the gap, and the radius is not much larger than the lattice constant.

\begin{figure}
    \centering
    \includegraphics[width=0.45\textwidth]{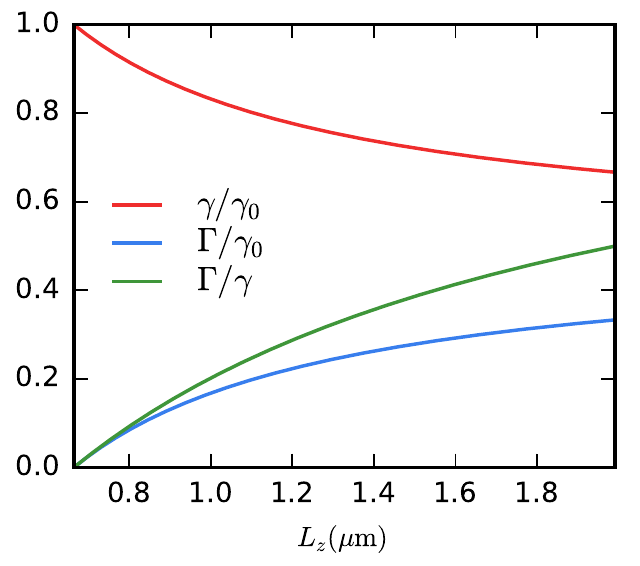}
    \caption{
        Actual strength of the couplings at $q$ such that $E_\ph(q) = E_\ex(q)$ as a function of $L_z$.
        The plotted range covers all $L_z$ values such that only one photon strongly interacts with the exciton.
}
    \label{fig:gammas}
\end{figure}

At this point, it is convenient to switch to the basis of circularly polarized photons.
That is the natural way to discuss the remarkable valley-polarization selection rule present in $\mathrm{MoS}_2$, which has been ubiquitously reported in the literature \cite{zeng2012valley,mak2012control}.
It is also advantageous regarding the topological description of photons, since the Berry curvature takes the form of a diagonal tensor in that basis \cite{onoda2004hall}.
Finally, the parameters $\Gamma_{\ex}$ and $\cos\theta$, which characterize the exciton and encode the cavity width, respectively, are recast in such a manner that their influence on the spectrum, selection rules and topology can be disclosed much more clearly.

For such a purpose, considering only photons with $k_z = \pi/L_z$ and using the notation defined in the main text, we introduce the operators $a_{\vec{q}}^{\pm}$ respective to circularly polarized light with 
\begin{eqnarray}
a_{\vec{q}}^{\rm TE}&=&\frac{i}{\sqrt 2}(e^{i\varphi}a_{\vec{q}}^+-e^{-i\varphi}a_{\vec{q}}^-)\, ,\\ \label{ate}
a_{\vec{q}}^{\rm TM}&=&\frac{1}{\sqrt 2}(e^{i\varphi}a_{\vec{q}}^+ + e^{-i\varphi}a_{\vec{q}}^-) \, . \label{atm}
\end{eqnarray}
They lead to Eq.
\eqref{eq:h_ph_ex}, which is one of the main results of our article.
We emphasize the fact that it has been derived microscopically, in contrast to the more heuristic approach of previous works \cite{karzig2015topological}.
As a consequence, the coupling strengths in the Hamiltonian, the sources of the winding phases $\varphi$---which encode the topological behavior---and their tunability are thoroughly clarified throughout our analysis. 

Notice that here the transformation between TE-TM and circularly polarized modes is performed through a canonical transformations that preserves the commutations relations, and therefore the statistics, of photons. 
A non-canonical transformation, such as the one used in Ref.~\cite{karzig2015topological}, leads to a photon mode whose energy goes to zero at large $q$ and an underestimation of the exciton-photon coupling in the relevant strong coupling regime ($q\simeq k_z$).

\begin{figure}
    \centering
    \includegraphics[width=0.45\textwidth]{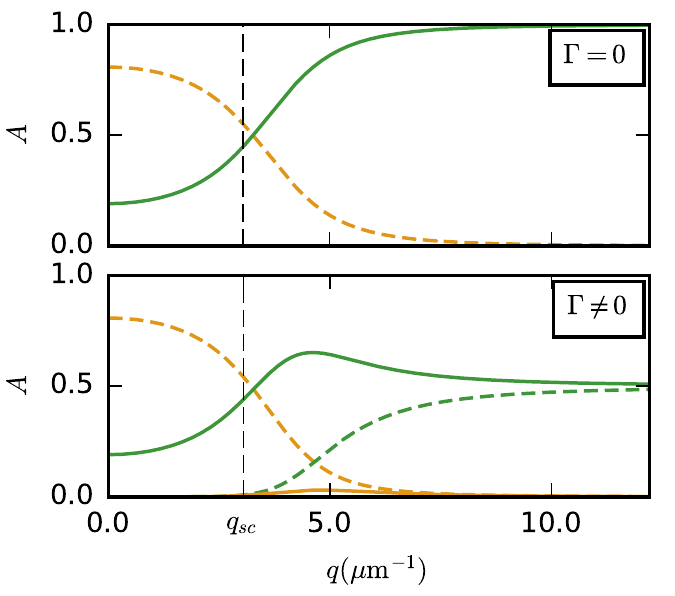}
    \caption{
        Probability amplitudes $A$ of the photonic (green) and the excitonic (orange) components for the upper polariton of highest energy.
    $\nu,\tau = +(-)$ correspond to solid (dashed) lines.
    The value of $q_{sc}$ is such that $E_{\ex}(q_{sc}) = E_{\ph}(q_{sc})$.
    }
    \label{fig:amplitudes}
\end{figure}

The great impact of the coupling $\Gamma$ on both the spectrum and the topological properties of the system is one of the main points of our work, and has been analyzed in detail along the main text.
Here, we provide with further comments to complete the discussion.
Fig.~\ref{fig:Fs} plots its magnitude and compares it to $\gamma$ in the range of cavity widths that our model properly describes, showing that it is far from being negligible in the most part of the domain.

Besides the splitting of UP and LP branches, $\Gamma$ deeply affects the nature of the hybrid modes.
Fig.~\ref{fig:amplitudes} illustrates this fact for UPs: at low momenta, there is little difference in the composition of polaritons with $\Gamma = 0$ and $\Gamma \neq 0$.
However, as long as $q$ goes beyond the strong-coupling regime, the former only involve photons with one circular polarization, whereas the latter evolve to a linearly polarized state of light.
This fact might be of relevance in polaritonic transport and its measurement via photonic polarization with similar techniques to those described in Refs.~\cite{onga2017exciton,liu2016strong}.

\section{Berry curvature of composite quasiparticles}

A generic single-particle Hamiltonian can be written as $H=\sum_{ij}[\hat{\phi}^i]^\dag{\cal H}_{ij}\hat{\phi}^j$, where $\hat{\phi}^j$ are second quantized operators describing the constituents. 
The operator $\hat{\phi}^j$ might depend on the parameter $\vec{q}$ involved in the calculation of the Berry connection and curvature. 
The wavefunction of the $n$th composite particle reads $\ket{ \psi^n } = \sum_j \psi_j^n [\hat{\phi}^j]^\dagger \ket{0}$, fulfilling ${\cal H}\psi_n=E_n \psi_n$, with $|0\rangle$ the vacuum. 

The Berry connection of the state $\ket{\psi^n}$ is defined as $\vec{A}^n = i\bra{\psi^n} \vec{\nabla}_{\vec{q}} \ket{\psi^n}$ and results in the form given in Eq.~\eqref{eq:berry_connection}
with the Berry connection of the bare constituents as $\vec{A}_{0}^{ij} = i \bra{0} \hat{\phi}^i \vec{\nabla}_{\vec{q}} [\hat{\phi}^j]^\dagger \ket{0}$.
The form of Eq.~\eqref{eq:berry_connection} is independent on the choice of the basis states defined by $\hat{\phi}^j$. 
It is straightforward to prove that upon a redefinition of the original modes by $\hat{\phi}^{i\prime} = {V^i}_l\hat{\phi}^{l}$, the expression for $\vec{A}^n$ is only affected by the substitution $\hat{\phi}^j \to \hat{\phi}^{j\prime}$ and $\psi_n\to \psi'_n=V^\dag \psi_n$. 
The Berry curvature can then be obtained as $\vec{\Omega}^n=\nabla_\vec{q}\times\vec{A}^n$.

The natural generalization of Eq.~(\ref{eq:berry_connection}) to situations where the state of the composite quasiparticle is degenerate, and non Abelian fields arise, is  $\vec{A}^{n,n'} = i[\psi^n]^\dag \nabla_\vec{q} \psi^{n'} + [\psi^n]^\dag \vec{A}_{0} \psi^{n'}$, with $\vec{\Omega}=\nabla_\vec{q}\times\vec{A}-i\vec{A}\times\vec{A}$.

\section{The origin of topology and effective models}

In the main text, we focused on the analysis of the BC for several cases $V_z, \Gamma \neq 0$, which is the most desirable in a prospective experiment.
Along this section, we provide with details of the effective model introduced in the text and discuss other situations to give a better understanding of the topological behavior and its enhancement by means of light-matter coupling.

All effective models presented here are obtained by applying a straightforward Schrieffer-Wolff transformation \cite{bravyi2011schrieffer-wolff,coleman2015introduction} to the subspaces spanned by either LPs or UPs separately.
This procedure yields 2-band models for photons or excitons at $q\to 0$, respectively, but including light-matter interactions and therefore accounting for the EBC.

For the case $V_z=0$ and at lowest order in $\gamma$ and $\Gamma$ we can derive the effective Hamiltonian
\begin{equation}\label{H2x2}
    H^s_{\rm eff} =
    \tilde{E}_s +
    \frac{\gamma \tilde\Gamma}{E} 
    \left(
        \begin{array}{cc}
            0               & q^2e^{2i\phi}     \\
            q^2e^{-2i\phi}  & 0
        \end{array}
    \right) \, ,
\end{equation}
with $E=\sqrt{\epsilon^2+\gamma^2}$, $\epsilon=(E_{\ex}-E_{\ph})/2$, $\tilde{E}_s=(E_{\ph}+E_{\ex})/2+s(E+\frac{\Gamma^2\epsilon^2}{2E^3})$ and $\Gamma=\tilde{\Gamma}q^2$.
Here, $s=1(-1)$ refers to UP (LP).
We point out that the Hamiltonian is time-reversal invariant, and up to the $q$-dependent global energy $\tilde{E}_s$, it is also particle-hole symmetric.
We then immediately find the BC of the split branches labeled by the index $\lambda = \pm$, namely 
\begin{equation}
\Omega_\pol^{s,\lambda}(\vec{q}) = 2\pi \lambda \delta^{(2)}(\vec{q}).
\end{equation}

\begin{figure}
    \centering
    \includegraphics[width=0.45\textwidth]{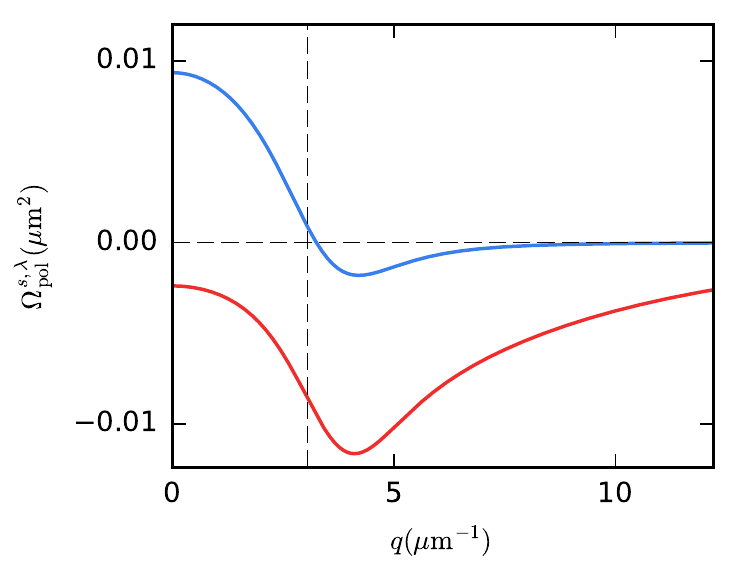}
    \caption{Same as in Fig.~\ref{fig:bc} but for $\Gamma = 0$, which results in a zero EBC.
    The difference in the scales of both plots should be noted.}
    \label{fig:bc_noGamma}
\end{figure}

For $V_z \neq 0$ a gap is opened in the effective Hamiltonian.
By treating $\gamma$ and $\Gamma$ as perturbative parameters and considering $V_z \ll |E_{\mathrm{ph}}-E_{\mathrm{ex}}|$ we find the Hamiltonian given in the text 
\begin{equation}
    H^s_{\rm eff} = 
    f_s(q) +
    \left(
        \begin{array}{cc}
            \Delta_s(q)                 &
            \alpha q^2e^{2i\phi}      \\
            \alpha^* q^2 e^{-2i\phi}  &
            -\Delta_s(q)
        \end{array}
    \right) \, ,
\end{equation}
where the parameters are
\begin{align*}
    & \alpha =
    -2\gamma\tilde{\Gamma} / \epsilon \, , \\
    & \Delta_{1}(q) = 
    V_z[1-(\gamma^2
    +\tilde{\Gamma}^2 q^4)/(2\epsilon^2)] \,,   \\
    & \Delta_{-1}(q) = 
    V_z(\gamma^2
    -\tilde{\Gamma}^2 q^4)/(2\epsilon^2) \,,    \\
    & f_s(q) = 
    E_s^{(0)}
    + s(\gamma^2+\tilde{\Gamma}^2 q^4) 
    /\epsilon \,, \\
    & E_{+1}^{(0)} = E_{\mathrm{ex}} \, , \ 
    E_{-1}^{(0)} = E_{\mathrm{ph}} \, .
\end{align*}
Interestingly, the gap $\Delta_s$ changes for the upper and lower branches, accounting for the difference in the topological behavior that is discussed in the main text.

We emphasize that in contrast to the previous case and as a result of time-reversal-symmetry breaking, the contribution of the intrinsic Berry curvature of the photons and the exciton does not cancel.
See Fig.
\ref{fig:bc} and the corresponding remarks for a thorough discussion.
This is an interesting point that nobody has considered yet.

Finally, we include in Fig.~\ref{fig:bc_noGamma} a plot for $\Gamma = 0$, $V_z \neq 0$ that can be directly compared to Fig.~\ref{fig:bc}.
We close this discussion by noting the striking differences between both cases.
The BC for $\Gamma = 0$ only comes from the intrinsic terms, whereas that of $\Gamma \neq 0$ has a more involved structure.
Not only does $\Gamma$ induce a qualitative change in $\Omega_n$, but it also boosts in one order of magnitude the maximum value of the BC.
This is reflected in the dynamics of polariton, sharpening the Hall effect with respect to bare excitons and photons, as we prove in the section about the polariton Hall effect.
In sum, we can conclude that light-matter coupling emerges as the main mechanism to increase the topology of hybrid modes high beyond that of their constituents.

\end{document}